\documentclass[prd,twocolumn,floatfix,amsmath,nofootinbib,amssymb,floatfix]{revtex4}
\usepackage{graphicx,color,dcolumn,booktabs,bm,multirow}
\usepackage{longtable,lscape}
\usepackage{txfonts}
\usepackage{overpic}
\usepackage{amssymb}
\usepackage{array}
\usepackage{indentfirst}
\usepackage{feynmf}   %{feynmp}
\usepackage{slashed}  %for Feynman symbols
\usepackage{setspace}
\usepackage{cases}
\usepackage{color}
\usepackage{multirow}
\usepackage{epstopdf}
\usepackage[utf8]{inputenc}
\usepackage{graphicx,color,dcolumn,booktabs,bm}
\usepackage[colorlinks,
citecolor=blue,
anchorcolor=red,
menucolor=red,
linkcolor=red,
filecolor=red,
runcolor=red,
urlcolor=blue,
frenchlinks=red]{hyperref}
\usepackage{float}

\begin{document}
\title{$\chi_{c0}$ production in  $\bar{\Lambda}n$ scattering process}
\author{Quan-Yun Guo$^{1}$}%\email{guoquanyun@seu.edu.cn}
\author{Zhuo Yu$^{1}$}
\author{Dian-Yong Chen$^{1,2}$\footnote{Corresponding author}}\email{chendy@seu.edu.cn}
\affiliation{$^1$ School of Physics, Southeast University, Nanjing 210094, People's Republic of China}
\affiliation{$^2$ Lanzhou Center for Theoretical Physics, Lanzhou University, Lanzhou 730000, China}
\date{\today}

	\begin{abstract}
In the present work, we investigate the production of $\chi_{c0}$ in the $\bar{\Lambda} n \rightarrow \chi_{c0} K^{0}_{S}$ process by utilizing an effective Lagrangian approach. Based on the $\chi_{c0} \rightarrow n K^{0}_{S} \bar{\Lambda}$ decay observed by the BESIII Collaboration, the coupling constants are determined by fitting the $M_{K^{0}_{S} \bar{\Lambda}}$ and $M_{n K^{0}_{S}}$ invariant mass spectra, where eight $N$ and $\Lambda$ state are considered. Moreover, the parameters $\beta$ is taken to be 1, 2, 3 as special cases to check the parameter dependence. Our results show that two clear structures can be observed in both $M_{K^{0}_{S} \bar{\Lambda}}$ and $M_{n K^{0}_{S}}$ invariant mass spectra, which should correspond to $N(1720)/N(1895)$ and $\Lambda(1600)/\Lambda(1890)$, respectively. In addition, the contributions from $N(938)$ and $\Lambda(1116)$ intermediate states are much smaller than those of the above-threshold resonances. With the fitted parameters for $\beta=1$, 2, 3, the cross sections for $\bar{\Lambda} n \rightarrow \chi_{c0} K^{0}_{S}$ are estimated to be 1.9 nb, 4.2 nb, and 8.0 nb at $\sqrt{s}=6$ GeV, respectively. Furthermore, our estimations indicate that the differential cross sections for $\bar{\Lambda} n \rightarrow \chi_{c0} K^{0}_{S}$ reach a maximum in the forward angle region.
	\end{abstract}

	\maketitle
	%\end{CJK}

\section{Introduction}
\label{sec:Introduction}
The (anti-)hyperon-nucleon ($YN/\bar{Y}N$) interaction is essential for understanding the strong interaction in the strange quark sector. Compared with the (anti-)nucleon-nucleon ($NN/\bar{N}N$) case~\cite{Machleidt:1987hj, Kaiser:1997mw, Kaplan:1996xu, Nakayama:1988zz, Williams:1987yn}, the (anti-)hyperon in $YN/\bar{Y}N$ scattering carries the strange quark, offering a unique probe of SU(3) flavor symmetry and its breaking in (anti-)baryon-baryon systems. However, precise theoretical descriptions of $YN/\bar{Y}N$ interactions remain challenging due to the limited experimental data and the non-perturbative nature of low-energy QCD, which motivates further investigations.

Experimentally, various hyperon-nucleon interactions have been studied since the 1960s, with $\Lambda p$ scattering being the most extensively investigated. These experiments cover a wide range of incident $\Lambda$ beam energies, including the low-energy range of 100-300 MeV~\cite{Alexander:1966ara, Alexander:1968acu, Sechi-Zorn:1968mao}, the intermediate energy range up to 5 GeV~\cite{Bassano:1967kbh, Alexander:1969tv, Charlton:1970bv, Hauptman:1977hr}, and the high-energy range up to about 20 GeV~\cite{Kubis:1970im, Anderson:1975rh, Mount:1975pb, Giacomelli:1976gn}. Unfortunately, progress in these areas stagnated in the following decades, primarily due to the instability of hyperon beams and limitations in detector resolution. A turning point came in 2021, when the CLAS Collaboration measured the cross sections for $\Lambda p \rightarrow \Lambda p$ with momentum  in the range $P_{\Lambda}=[0.9,2.0]$ GeV~\cite{CLAS:2021gur}. In 2024, using the $\Lambda /\bar{\Lambda}$ beam produced in the $J/\psi \to \Lambda \bar{\Lambda}$ decay, the BESIII Collaboration measured the $\Lambda p \rightarrow \Lambda p$ and $\bar{\Lambda} p \rightarrow \bar{\Lambda} p$ elastic scattering, and the relevant cross sections were reported to be $(12.2 \pm 1.6 \pm 1.1)$ mb and $(17.5 \pm 2.1 \pm 1.6)$ mb at $P_{\Lambda/\bar{\Lambda}}=1.074 \pm 0.017$ GeV~\cite{BESIII:2024geh}, respectively. In addition, a strong forward peak was observed in the differential cross sections for the $\bar{\Lambda} p \rightarrow \bar{\Lambda} p$ process. Beyond the $\Lambda p$ and $\bar{\Lambda} p$ scattering, other (anti-)hyperon-nucleon interactions have also been explored experimentally, including $\Sigma^{+} n \rightarrow \Lambda p$ and $\Sigma^{+} n \rightarrow \Sigma^{0} p$~\cite{BESIII:2025bft}, $\Xi^{0}n \rightarrow \Lambda \Lambda$ and $\Xi^{0}n \rightarrow \Lambda \Xi^{0}$~\cite{BESIII:2025vqk}.

Theoretically, a variety of approaches have been employed to investigate the $\Lambda p$ interaction. In Ref.~\cite{Holzenkamp:1989tq}, the authors estimated the cross sections for $\Lambda N \rightarrow \Lambda N$ using the meson exchange model constrained by experimental data, which incorporates both single-meson exchange and higher-order processes involving spin-$3/2$ baryon exchange. Moreover, the authors in Refs.~\cite{Rijken:1998yy, Tominaga:2001ra, Haidenbauer:2005zh} studied the hyperon-nucleon interactions using the one-boson-exchange model, in which the cross sections and differential cross sections for $\Lambda p \rightarrow \Lambda p$ were investigated. The chiral effective field theory~\cite{Polinder:2006zh, Haidenbauer:2013oca, Li:2016mln, Haidenbauer:2019boi, Song:2021yab, Haidenbauer:2023qhf} has also been applied to study the $\Lambda p$ interaction, with calculations ranging from leading order to next-to-leading order. Using an effective Lagrangian approach, the authors in Ref.~\cite{Wang:2024whi} estimated the cross sections for the $\Lambda p \rightarrow \Lambda p$ and $\bar{\Lambda} p \rightarrow \bar{\Lambda} p$ processes, with the former involving $u$-channel $K, K^{\ast}$ exchanges and $t$-channel $\eta, \omega, \sigma$ exchanges, while the latter involves only $t$-channel $\eta, \omega, \sigma$ exchanges. Within this framework, the measured cross sections and differential cross sections reported by the BESIII Collaborations~\cite{BESIII:2024geh} can be well reproduced~\cite{Wang:2024whi}. Following the same approach, the authors in Ref.~\cite{Chen:2025qxh} investigated the same processes within the quasipotential Bethe-Salpeter equation framework, where the relevant cross sections and differential cross sections were also estimated. More recently, the authors in Ref.~\cite{Liu:2026gxr} investigated the $p-\Lambda$ scattering using  lattice QCD with the pion masses ranging from 135.5 MeV to 317.2 MeV. Their estimations indicated that the $p-\Lambda$ system exhibits weak attractive interactions within the considered energy regime. In addition, the scattering length and effective range were determined to be 0.177(83) GeV and 2.9(1.4) fm for the $^{1}S_{0}$ channel, and 0.016(76) GeV and 1.8(1.1) fm for the $^{3}S_{1}$ channel, respectively.

Notably, charmonium decays offer a unique avenue for investigating (anti-)hyperon-nucleon scattering, for example, the $\Lambda p/\bar{\Lambda}p$ elastic scattering~\cite{BESIII:2024geh}, $\Sigma^+ n/\Xi^0n$ inelastic scattering~\cite{BESIII:2025bft,BESIII:2025vqk}. Besides the above measured processes, charmonium production in baryon-antibaryon scattering processes have been investigated by connecting them with the charmonium decay processes via crossing symmetry, for example, the $p\bar{p} \to J/\psi\pi^0$~\cite{Lundborg:2005am},  $p\bar{p} \to \psi(3770) \pi^0$~\cite{Xu:2015qqa}, $p\bar{p} \to Y(4220) \pi^0$~\cite{Wang:2017sxq}, $p\bar{p}\to J/\psi \eta$~\cite{Zhou:2020smj}. In 2021, the BESIII Collaboration studied the $\chi_{cJ} \rightarrow n K^{0}_{S} \bar{\Lambda}$ $(J=0, 1, 2)$ decays by analyzing $(4.48 \pm 0.03) \times 10^{8}$ $\psi(3686)$ events collected with the BESIII detector~\cite{BESIII:2021haw}, where the $M_{n \bar{\Lambda}}$, $M_{n K^{0}_{S}}$, and $M_{K^{0}_{S} \bar{\Lambda}}$ invariant mass spectra of the $\chi_{cJ}$ decays were presented. The branching fractions were measured to be $(6.67\pm 0.26 \pm 0.41)\times 10^{-4}$, $(1.71\pm 0.12 \pm 0.12) \times 10^{-4}$, and $(3.66\pm 0.17 \pm 0.23) \times 10^{-4}$ for $J=0$, $1$, and $2$, respectively. Benefitting from the large branching fraction, precise $M_{n \bar{\Lambda}}$, $M_{n K^{0}_{S}}$, and $M_{K^{0}_{S} \bar{\Lambda}}$ invariant mass distributions of $\chi_{c0}\to nK_S^0 \bar{\Lambda}$ have been reported by the BESIII Collaboration~\cite{BESIII:2021haw}. As indicated in Fig.~\ref{Fig.1}, $\bar{N}$ and $\Lambda$ resonances can serve as intermediate states of the $\chi_{c0} \rightarrow n K^{0}_{S} \bar{\Lambda}$ process, with their contributions manifesting as structures in the $M_{K^{0}_{S} \bar{\Lambda}}$ and $M_{n K^{0}_{S}}$ invariant mass spectra. Then the relevant coupling strength can be well determined by fitting the measured $M_{K^{0}_{S} \bar{\Lambda}}$ and $M_{n K^{0}_{S}}$ invariant mass spectra. Consequently, we can investigate the process $\bar{\Lambda} n \to \chi_{c0} K_S^0$ with these determined coupling strength, where the $u$- and $t$-channel exchanges by $N$ and $\Lambda$ resonances can be taken into considerations.

This work is organized as follows. After the introduction, the effective Lagrangian approach used to investigate the $\chi_{c0} \rightarrow n K^{0}_{S} \bar{\Lambda}$ decay and $\bar{\Lambda} n \rightarrow \chi_{c0} K^{0}_{S}$ scattering is presented. In Sec. \ref{sec:MA}, the model parameters are obtained by fitting the $M_{K^{0}_{S} \bar{\Lambda}}$ and $M_{n K^{0}_{S}}$ invariant mass spectra of the $\chi_{c0}$ decay. Then, the numerical results and relevant discussions of the cross sections for $\bar{\Lambda} n \rightarrow \chi_{c0} K^{0}_{S}$ are presented. The last section is devoted to a summary.

\section{Theoretical framework for the $\chi_{c0}$ decay and $\bar{\Lambda} n$ scattering}
\label{sec:MS}
\subsection{The $\chi_{c0} \rightarrow n K^{0}_{S} \bar{\Lambda}$ decay}
%%%%%%%%%%%%%%%%%%%%%%%%%%%%%%%%%%%%%%%%%%%%%%%%%%%
\begin{figure}[t]
		\centering
		\includegraphics[width=75mm]{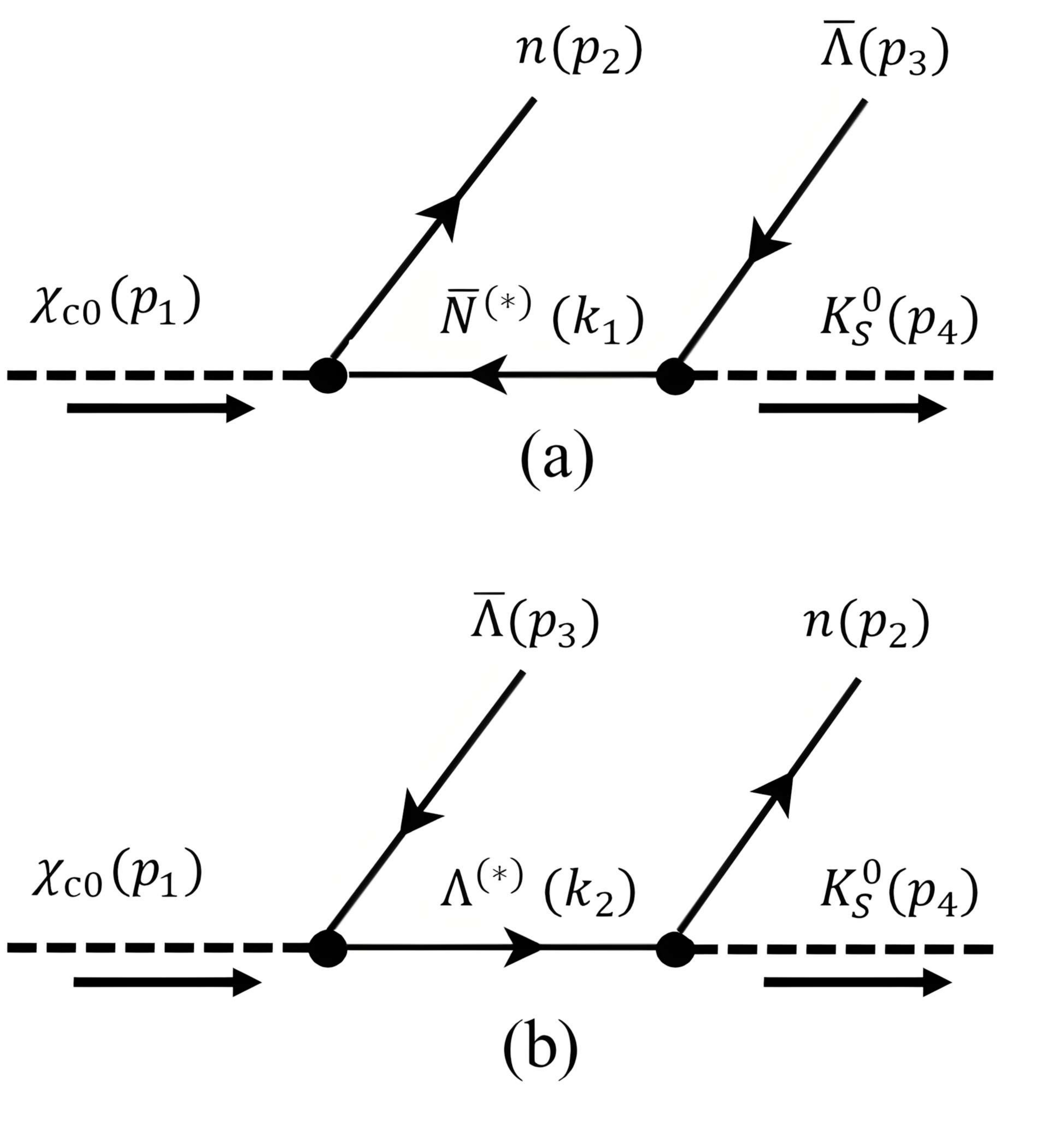}
	\caption{Diagrams contributing to the $\chi_{c0} \rightarrow n K^{0}_{S} \bar{\Lambda}$ decay. Diagrams (a) and (b) correspond to the $\bar{N}^{(\ast)}$ and $\Lambda^{(\ast)}$ intermediate processes, respectively. Here $\bar{N}^{(\ast)}$ and $\Lambda^{(\ast)}$ refer to $\bar{N}$ and $\Lambda$ baryons and their resonances, respectively.} \label{Fig.1}
\end{figure}
%%%%%%%%%%%%%%%%%%%%%%%%%%%%%%%%%%%%%%%%%%%%%%%%%%
In 2021, the BESIII Collaboration observed the $\chi_{c0} \rightarrow n K^{0}_{S} \bar{\Lambda}$ decay and reported the $M_{n \bar{\Lambda}}$, $M_{n K^{0}_{S}}$, and $M_{K^{0}_{S} \bar{\Lambda}}$ invariant mass spectra as well~\cite{BESIII:2021haw}. Among them, the structures observed in the $M_{K^{0}_{S} \bar{\Lambda}}$ and $M_{n K^{0}_{S}}$ invariant mass spectra can be attributed to the contributions from $\bar{N}$ and $\Lambda$ baryons and their resonances. Motivated by these experimental observations, we employ the effective Lagrangian approach to depict the relevant hadron interaction vertices in the present work. The diagrams contributing to $\chi_{c0} \rightarrow n K^{0}_{S} \bar{\Lambda}$ are listed in Fig.~\ref{Fig.1}, where $\bar{N}^{(\ast)}$ and $\Lambda^{(\ast)}$ refer to the $\bar{N}$ and $\Lambda$ baryons and their resonances, respectively. In the following, we assign the notation $B$ to represent the neutron and $\bar{\Lambda}$ final states, and $Y_{J^{P}}$ to represent the $\bar{N}^{(\ast)}$ and $\Lambda^{(\ast)}$ intermediate states with $J^{P}$ quantum numbers. Moreover, we only take  lower-spin states with $J\leq 3/2$ into account, as these states are expected to dominate the physical processes near the threshold region. The effective Lagrangians for $B Y_{1/2^{\pm}} \chi_{c0}$, $B Y_{3/2^{\pm}} \chi_{c0}$, $B Y_{1/2^{\pm}} K^{0}_{S}$, and $B Y_{3/2^{\pm}} K^{0}_{S}$ read~\cite{Chen:2011xk,Okubo:1975sc,Jackson:2015dva,Oh:2000qr,Tsushima:1996xc,Tsushima:1996tv,Tsushima:1998jz,Ouyang:2009kv},
\begin{eqnarray}
\mathcal{L}_{B Y_{1/2^{\pm}} \chi_{c0}}&=& f_{B Y_{1/2^{\pm}} \chi_{c0}} \bar{B} \left\{
\begin{array}{c}
	1 \\
	\gamma_5
\end{array}	
\right\} Y_{1/2^{\pm}} \chi_{c0} +H.c.,\nonumber\\ 
%\end{eqnarray}
%\begin{eqnarray}
\mathcal{L}_{B Y_{3/2^{\pm}} \chi_{c0}}&=& \frac{f_{B Y_{3/2^{\pm}} \chi_{c0}}}{m_{\pi}} \bar{B} \left\{
\begin{array}{c}
	\gamma_5 \\
	1
\end{array}	
\right\} Y^{\mu}_{3/2^{\pm}} \partial_{\mu} \chi_{c0} +H.c.,\nonumber\\ 
%\end{eqnarray}
%\begin{eqnarray}
\mathcal{L}_{B Y_{1/2^{\pm}} K^{0}_{S}}&=& (\mp i) f_{B Y_{1/2^{\pm}} K^{0}_{S}} \bar{B} \left\{
\begin{array}{c}
	\gamma_5 \\
	1
\end{array}	
\right\} Y_{1/2^{\pm}} K^{0}_{S} +H.c.,\nonumber\\ 
%\end{eqnarray}
%\begin{eqnarray}
\mathcal{L}_{B Y_{3/2^{\pm}} K^{0}_{S}}&=& \frac{f_{B Y_{3/2^{\pm}} K^{0}_{S}}}{m_{\pi}} \bar{B} \left\{
\begin{array}{c}
	1 \\
	\gamma_5
\end{array}	
\right\} Y^{\mu}_{3/2^{\pm}} \partial_{\mu} K^{0}_{S} +H.c., \label{Eq.1}
\end{eqnarray}
where the upper and lower symbols in the curly brackets refer to the positive and negative parity of the $\bar{N}^{(\ast)}$ and $\Lambda^{(\ast)}$ states, respectively. With these effective Lagrangians, one can obtain the amplitudes of the $\chi_{c0} \rightarrow n K^{0}_{S} \bar{\Lambda}$ decay, which are,
\begin{eqnarray}
\mathcal{M}_{N_{1/2^{+}}}&=&\bar{u}(p_{2}) \Big[f_{n N_{1/2^{+}} \chi_{c0}}\Big] \mathcal{S}^{1/2}(k_{1}, m_{N}, \Gamma_{N}) \nonumber\\ &\times& \Big[-i f_{\Lambda N_{1/2^{+}} K^{0}_{S}} \gamma_{5}\Big] v(p_{3}) \Big[F(k_{1}, m_{N}, \Lambda_{N}) \Big]^2,\nonumber\\
%\end{eqnarray}
%\begin{eqnarray}
\mathcal{M}_{\Lambda_{1/2^{+}}}&=&\bar{u}(p_{2}) \Big[-i f_{n \Lambda_{1/2^{+}} K^{0}_{S}} \gamma_{5}\Big] \mathcal{S}^{1/2}(k_{2}, m_{\Lambda}, \Gamma_{\Lambda}) \nonumber\\ &\times& \Big[f_{\Lambda \Lambda_{1/2^{+}} \chi_{c0}}\Big] v(p_{3}) \Big[F(k_{2}, m_{\Lambda}, \Lambda_{\Lambda}) \Big]^2,\nonumber\\
%\end{eqnarray}
%\begin{eqnarray}
\mathcal{M}_{N_{1/2^{-}}}&=&\bar{u}(p_{2}) \Big[f_{n N_{1/2^{-}} \chi_{c0}} \gamma_{5} \Big] \mathcal{S}^{1/2}(k_{1}, m_{N}, \Gamma_{N}) \nonumber\\ &\times& \Big[i f_{\Lambda N_{1/2^{-}} K^{0}_{S}} \Big] v(p_{3}) \Big[F(k_{1}, m_{N}, \Lambda_{N}) \Big]^2,\nonumber\\
%\end{eqnarray}
%\begin{eqnarray}
\mathcal{M}_{\Lambda_{1/2^{-}}}&=&\bar{u}(p_{2}) \Big[i f_{n \Lambda_{1/2^{-}} K^{0}_{S}} \Big] \mathcal{S}^{1/2}(k_{2}, m_{\Lambda}, \Gamma_{\Lambda}) \nonumber\\ &\times& \Big[f_{\Lambda \Lambda_{1/2^{-}} \chi_{c0}} \gamma_{5} \Big] v(p_{3}) \Big[F(k_{2}, m_{\Lambda}, \Lambda_{\Lambda}) \Big]^2,\nonumber
\end{eqnarray}
\begin{eqnarray}
\mathcal{M}_{N_{3/2^{+}}}&=&\bar{u}(p_{2}) \Big[\frac{f_{n N_{3/2^{+}} \chi_{c0}}}{{ m_{\pi}}} \gamma_{5} (-i p^{\mu}_{1}) \Big] \mathcal{S}^{3/2}_{\mu \nu}(k_{1}, m_{N}, \Gamma_{N}) \nonumber\\ &\times& \Big[\frac{f_{\Lambda N_{3/2^{+}} K^{0}_{S}}}{{ m_{\pi}}} (i p^{\nu}_{4}) \Big] v(p_{3}) \Big[F(k_{1}, m_{N}, \Lambda_{N}) \Big]^2,\nonumber\\
%\end{eqnarray}
%\begin{eqnarray}
\mathcal{M}_{\Lambda_{3/2^{+}}}&=&\bar{u}(p_{2}) \Big[\frac{f_{n \Lambda_{3/2^{+}} K^{0}_{S}}}{{ m_{\pi}}} (i p^{\nu}_{4}) \Big] \mathcal{S}^{3/2}_{\mu \nu}(k_{2}, m_{\Lambda}, \Gamma_{\Lambda}) \nonumber\\ &\times& \Big[\frac{f_{\Lambda \Lambda_{3/2^{+}} \chi_{c0}}}{{ m_{\pi}}} \gamma_{5} (-i p^{\mu}_{1}) \Big] v(p_{3}) \Big[F(k_{2}, m_{\Lambda}, \Lambda_{\Lambda}) \Big]^2,\nonumber\\
%\end{eqnarray}
%\begin{eqnarray}
\mathcal{M}_{N_{3/2^{-}}}&=&\bar{u}(p_{2}) \Big[\frac{f_{n N_{3/2^{-}} \chi_{c0}}}{{ m_{\pi}}} (-i p^{\mu}_{1}) \Big] \mathcal{S}^{3/2}_{\mu \nu}(k_{1}, m_{N}, \Gamma_{N}) \nonumber\\ &\times& \Big[\frac{f_{\Lambda N_{3/2^{-}} K^{0}_{S}}}{{ m_{\pi}}} \gamma_{5} (i p^{\nu}_{4}) \Big] v(p_{3}) \Big[F(k_{1}, m_{N}, \Lambda_{N}) \Big]^2,\nonumber\\
%\end{eqnarray}
%\begin{eqnarray}
\mathcal{M}_{\Lambda_{3/2^{-}}}&=&\bar{u}(p_{2}) \Big[\frac{f_{n \Lambda_{3/2^{-}} K^{0}_{S}}}{{ m_{\pi}}} \gamma_{5} (i p^{\nu}_{4}) \Big] \mathcal{S}^{3/2}_{\mu \nu}(k_{2}, m_{\Lambda}, \Gamma_{\Lambda}) \nonumber\\ &\times& \Big[\frac{f_{\Lambda \Lambda_{3/2^{-}} \chi_{c0}}}{{ m_{\pi}}} (-i p^{\mu}_{1}) \Big] v(p_{3}) \Big[F(k_{2}, m_{\Lambda}, \Lambda_{\Lambda}) \Big]^2,
\label{Eq.2} 
\end{eqnarray}
where the subscripts $N_{J^{P}}$ and $\Lambda_{J^{P}}$ correspond to the $\bar{N}^{(\ast)}$ and $\Lambda^{(\ast)}$ propagators with the corresponding $J^{P}$ quantum numbers. Moreover, in the above amplitudes, $\mathcal{S}^{1/2}(k_{i},m_{i},\Gamma_{i})$ and $\mathcal{S}^{3/2}_{\mu \nu}(k_{i},m_{i},\Gamma_{i})$ refer to the propagators of the baryons with spin $1/2$ and spin $3/2$, respectively, and their concrete forms are,
\begin{eqnarray}
&&\mathcal{S}^{1/2}(k_{i},m_{i},\Gamma_{i}) = \frac{\slash\!\!\!k_{i}+m_{i}} {k^2_{i}-m^2_{i} +i m_{i} \Gamma_{i}},\nonumber\\
%\end{eqnarray}
%\begin{eqnarray}
&&\mathcal{S}^{3/2}_{\mu \nu}(k_{i},m_{i},\Gamma_{i})  = \frac{\slash\!\!\!k_{i}+m_{i}} {k^2_{i}-m^2_{i} +i m_{i} \Gamma_{i}} \nonumber\\&&\qquad \qquad\times  \Big(-g^{\mu \nu} + \frac{\gamma^{\mu} \gamma^{\nu}}{3} + \frac{2 k^{\mu}_{i} k^{\nu}_{i}}{3 m^{2}_{i}} +\frac{\gamma^{\mu} k^{\nu}_{i}-k^{\mu}_{i} \gamma^{\nu}}{3 m_{i}} \Big),\qquad \label{Eq.3}
\end{eqnarray}
respectively.

In addition, the form factor $F(k_{i},m_{Y}, \Lambda_{Y})$ is introduced to depict the inner structure of the involved hadrons in each vertex, and its specific expression is,
\begin{eqnarray}
F(k_{i},m_{Y},\Lambda_{Y})
=\frac{\Lambda^{4}_{Y}} {\Lambda^{4} _{Y} +(k^{2}_{i}-m^{2}_{Y})^2}, \label{Eq.4}
\end{eqnarray}
where $k_{i}$ and $m_{Y}$ stand for the four-momentum and the mass of the exchanged state, respectively. The cutoff parameter $\Lambda_{Y}$ can be parameterized as,
\begin{eqnarray}
\Lambda_{Y}
=m_{Y}+\beta \Lambda_{QCD}, \label{Eq.5}
\end{eqnarray}
with $\Lambda_{QCD}=220$ MeV. In the present work, we take $\beta=1,2,$ and $3$ as special cases to check the parameter dependence of the considered processes.

%%%%%%%%%%%%%%%%%%%%%%%%%%%%%%%%%%%%%%%%%%%%%%%%%%%
\begin{figure}[htb]
		\centering
		\includegraphics[width=75mm]{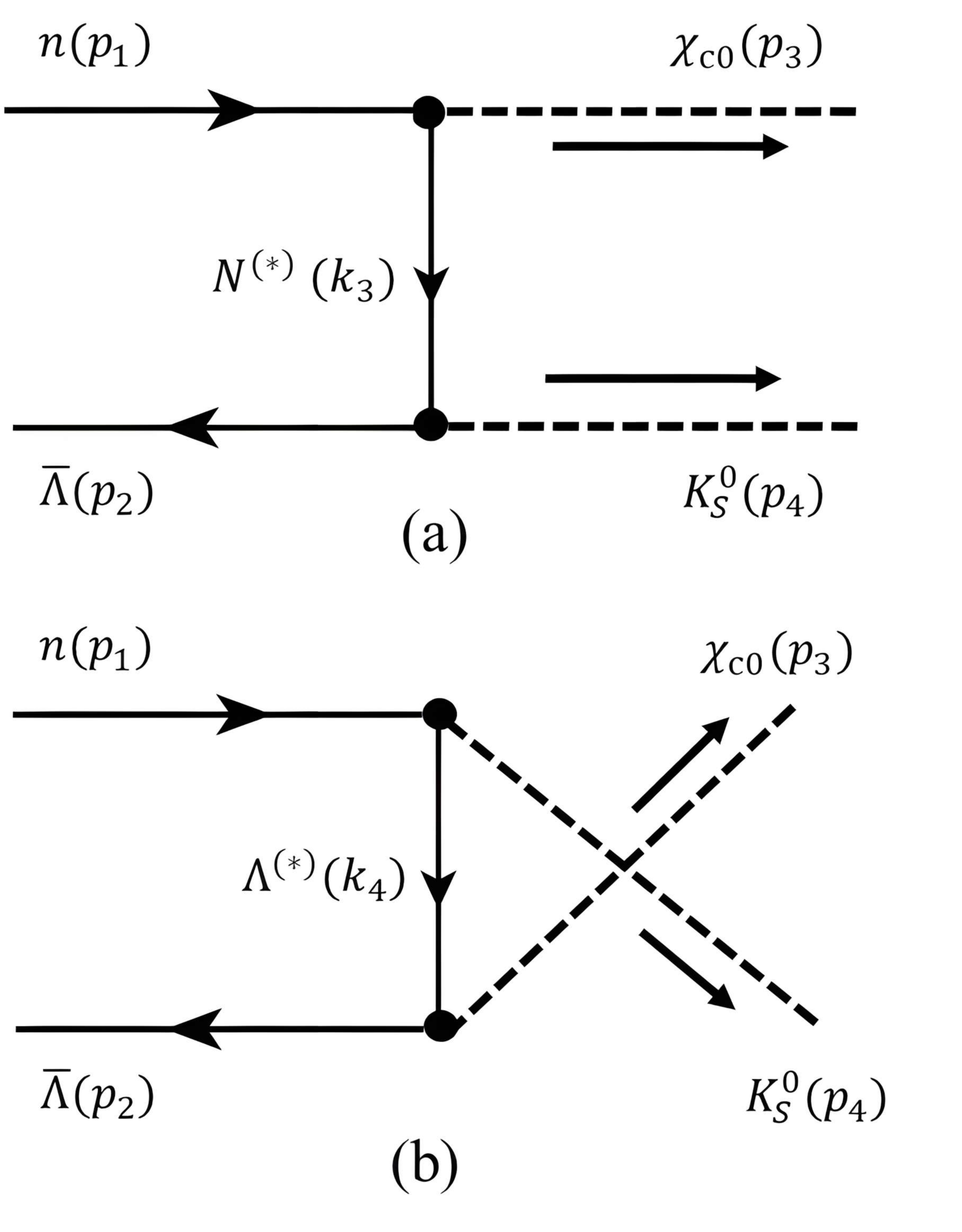}
	\caption{Diagrams contributing to the $\bar{\Lambda} n \rightarrow \chi_{c0} K^{0}_{S}$ process. Diagrams (a) and (b) correspond to the $t$- and $u$-channel contributions, respectively. Here $N^{(\ast)}$ and $\Lambda^{(\ast)}$ refer to $N$ and $\Lambda$ baryons and their resonances.} \label{Fig.2}
\end{figure}
%%%%%%%%%%%%%%%%%%%%%%%%%%%%%%%%%%%%%%%%%%%%%%%%%%
\subsection{The $\bar{\Lambda} n \rightarrow \chi_{c0} K^{0}_{S}$ scattering}
Motivated by the feasibility of hyperon-nucleon scattering and potential higher-energy configurations in future upgrades, we investigate the production of $\chi_{c0}$ in the $\bar{\Lambda} n \rightarrow \chi_{c0} K^{0}_{S}$ process in the following. The diagrams contributing to $\bar{\Lambda} n \rightarrow \chi_{c0} K^{0}_{S}$ are presented in Fig.~\ref{Fig.2}, where both the $t$-channel $N^{(\ast)}$ exchanges [diagram (a)] and $u$-channel $\Lambda^{(\ast)}$ exchanges [diagram (b)] are included. In addition, the intermediate states considered in the $\bar{\Lambda} n$ scattering are the same as those in the $\chi_{c0}$ decay. With the effective Lagrangians in Eq.~\eqref{Eq.1}, one can obtain the amplitudes of the $\bar{\Lambda} n \rightarrow \chi_{c0} K^{0}_{S}$ process, which are,
\begin{eqnarray}
\mathcal{M}^{t}_{N_{1/2^{+}}}&=&\bar{v}(p_{2}) \Big[-i f_{\Lambda N_{1/2^{+}} K^{0}_{S}} \gamma_{5} \Big] \mathcal{S}^{1/2}(k_{3}, m_{N}, \Gamma_{N}) \nonumber\\ &\times& \Big[f_{n N_{1/2^{+}} \chi_{c0}} \Big] u(p_{1}) \Big[F(k_{3}, m_{N}, \Lambda_{N}) \Big]^2,\nonumber\\
%\end{eqnarray}
%\begin{eqnarray}
\mathcal{M}^{u}_{\Lambda_{1/2^{+}}}&=&\bar{v}(p_{2}) \Big[f_{\Lambda \Lambda_{1/2^{+}} \chi_{c0}} \Big] \mathcal{S}^{1/2}(k_{4}, m_{\Lambda}, \Gamma_{\Lambda}) \nonumber\\ &\times& \Big[-i f_{n \Lambda_{1/2^{+}} K^{0}_{S}} \gamma_{5} \Big] u(p_{1}) \Big[F(k_{4}, m_{\Lambda}, \Lambda_{\Lambda}) \Big]^2,\nonumber\\
%\end{eqnarray}
%\begin{eqnarray}
\mathcal{M}^{t}_{N_{1/2^{-}}}&=&\bar{v}(p_{2}) \Big[i f_{\Lambda N_{1/2^{-}} K^{0}_{S}} \Big] \mathcal{S}^{1/2}(k_{3}, m_{N}, \Gamma_{N}) \nonumber\\ &\times& \Big[f_{n N_{1/2^{-}} \chi_{c0}} \gamma_{5} \Big] u(p_{1}) \Big[F(k_{3}, m_{N}, \Lambda_{N}) \Big]^2,\nonumber\\
%\end{eqnarray}
%\begin{eqnarray}
\mathcal{M}^{u}_{\Lambda_{1/2^{-}}}&=&\bar{v}(p_{2}) \Big[f_{\Lambda \Lambda_{1/2^{-}} \chi_{c0}} \gamma_{5} \Big] \mathcal{S}^{1/2}(k_{4}, m_{\Lambda}, \Gamma_{\Lambda}) \nonumber\\ &\times& \Big[i f_{n \Lambda_{1/2^{-}} K^{0}_{S}} \Big] u(p_{1}) \Big[F(k_{4}, m_{\Lambda}, \Lambda_{\Lambda}) \Big]^2,\nonumber\\
%\end{eqnarray}
%\begin{eqnarray}
\mathcal{M}^{t}_{N_{3/2^{+}}}&=&\bar{v}(p_{2}) \Big[\frac{f_{\Lambda N_{3/2^{+}} K^{0}_{S}}}{{ m_{\pi}}} (i p^{\mu}_{4}) \Big] \mathcal{S}^{3/2}_{\mu \nu}(k_{3}, m_{N}, \Gamma_{N}) \nonumber\\ &\times& \Big[\frac{f_{n N_{3/2^{+}} \chi_{c0}}}{{ m_{\pi}}} \gamma_{5} (i p^{\nu}_{3}) \Big] u(p_{1}) \Big[F(k_{3}, m_{N}, \Lambda_{N}) \Big]^2,\nonumber
\end{eqnarray}
\begin{eqnarray}
\mathcal{M}^{u}_{\Lambda_{3/2^{+}}}&=&\bar{v}(p_{2}) \Big[\frac{f_{\Lambda \Lambda_{3/2^{+}} \chi_{c0}}}{{ m_{\pi}}} \gamma_{5} (i p^{\nu}_{3}) \Big] \mathcal{S}^{3/2}_{\mu \nu}(k_{4}, m_{\Lambda}, \Gamma_{\Lambda}) \nonumber\\ &\times& \Big[\frac{f_{n \Lambda_{3/2^{+}} K^{0}_{S}}}{{ m_{\pi}}} (i p^{\mu}_{4}) \Big]  u(p_{1}) \Big[F(k_{4}, m_{\Lambda}, \Lambda_{\Lambda}) \Big]^2,\nonumber\\
%\end{eqnarray}
%\begin{eqnarray}
\mathcal{M}^{t}_{N_{3/2^{-}}}&=&\bar{v}(p_{2}) \Big[\frac{f_{\Lambda N_{3/2^{-}} K^{0}_{S}}}{{ m_{\pi}}} \gamma_{5} (i p^{\mu}_{4}) \Big] \mathcal{S}^{3/2}_{\mu \nu}(k_{3}, m_{N}, \Gamma_{N}) \nonumber\\ &\times& \Big[\frac{f_{n N_{3/2^{-}} \chi_{c0}}}{{ m_{\pi}}} (i p^{\nu}_{3}) \Big] u(p_{1}) \Big[F(k_{3}, m_{N}, \Lambda_{N}) \Big]^2, \nonumber\\
%\end{eqnarray}
%\begin{eqnarray}
\mathcal{M}^{u}_{\Lambda_{3/2^{-}}}&=&\bar{v}(p_{2}) \Big[\frac{f_{\Lambda \Lambda_{3/2^{-}} \chi_{c0}}}{{ m_{\pi}}} (i p^{\nu}_{3}) \Big]  \mathcal{S}^{3/2}_{\mu \nu}(k_{4}, m_{\Lambda}, \Gamma_{\Lambda}) \nonumber\\ &\times& \Big[\frac{f_{n \Lambda_{3/2^{-}} K^{0}_{S}}}{{ m_{\pi}}} \gamma_{5} (i p^{\mu}_{4}) \Big] u(p_{1}) \Big[F(k_{4}, m_{\Lambda}, \Lambda_{\Lambda}) \Big]^2, \label{Eq.6}
\end{eqnarray}
where the superscripts $t$ and $u$ correspond to $t$- and $u$-channels, respectively. The subscripts $N_{J^{P}}$ and $ \Lambda_{J^{P}}$ correspond to the propagators with the corresponding $J^{P}$ quantum numbers. Moreover, the details of the involved $N^{(\ast)}$ and $\Lambda^{(\ast)}$ states will be discussed in the following section.

%%%%%%%%%%%%%%%%%%%%%%%%%%%%%%%%%%%%%%%%%%%%%%%%%%
\begin{table}
	\caption{The resonance parameters of the involved $N^{(\ast)}$ and $\Lambda^{(\ast)}$ baryons are taken from the Particle Data
Group~\cite{ParticleDataGroup:2026mpi}.}
	\label{Tab.1}
\renewcommand{\arraystretch}{2.0}
	\setlength{\tabcolsep}{10pt}
	\centering
	\begin{tabular}{cccc}
		\toprule[0.5pt]
\hline\hline
		\toprule[0.5pt]
		State&$J^{P}$&Mass (MeV)& Width (MeV)\\
		\midrule[1pt]
		$N(938)$&$1/2^{+}$&938&--\\
		$N(1720)$&$3/2^{+}$&1720&250\\
		$N(1895)$&$1/2^{-}$&1895&120\\
		$N(2040)$&$3/2^{+}$&2040&230\\
		$\Lambda(1116)$&$1/2^{+}$&1116&--\\
		$\Lambda(1600)$&$1/2^{+}$&1600&200\\
		$\Lambda(1800)$&$1/2^{-}$&1800&200\\
		$\Lambda(1890)$&$3/2^{+}$&1890&120\\
		\toprule[0.5pt]
\hline\hline
		\toprule[0.5pt]
	\end{tabular}
\end{table}
%%%%%%%%%%%%%%%%%%%%%%%%%%%%%%%%%%%%%%%%%%%%%%%%%
%%%%%%%%%%%%%%%%%%%%%%
\begin{table}
	\caption{The fitted parameters of the $\chi_{c0} \rightarrow n K^{0}_{S} \bar{\Lambda}$ decay for $\beta=1$, 2, and 3, respectively.}
	\label{Tab.2}
	\renewcommand{\arraystretch}{2}
	\setlength{\tabcolsep}{9pt}
	\centering
	\begin{tabular}{cccc}
		\toprule[0.5pt]
\hline\hline
		\toprule[0.5pt]
		$g_{Y}$ $\times$ $(10^{-3})$& $\beta=1$ & $\beta=2$ & $\beta=3$\\
		\midrule[1pt]
           $g_{N(938)}$&$110^{+198}_{-198}$& 
           $90.7^{+60.3}_{-60.3}$&$91.7^{+114}_{-114}$\\
		$g_{N(1720)}$&$0.18^{+0.04}_{-0.04}$&
		$0.14^{+0.05}_{-0.05}$&$0.10^{+0.12}_{-0.12}$\\
		$g_{N(1895)}$&$6.80^{+1.89}_{-1.89}$&
		$6.71^{+1.44}_{-1.44}$&$ 6.77^{+2.56}_{-2.56}$\\
		$g_{N(2040)}$&$-0.36^{+0.08}_{-0.08}$&
		$-0.39^{+0.05}_{-0.05}$&$-0.42^{+0.08}_{-0.08}$\\
           $g_{\Lambda(1116)}$&$123^{+99.8}_{-99.8}$& 
           $129^{+44.5}_{-44.5}$&$143^{+79.9}_{-79.9}$\\
		$g_{\Lambda(1600)}$&$277^{+12.0}_{-12.0}$&
		$277^{+10.0}_{-10.0}$&$276^{+12.3}_{-12.3}$\\
		$g_{\Lambda(1800)}$&$-13.5^{+1.54}_{-1.54}$&
		$-13.1^{+2.22}_{-2.22}$&$ -12.4^{+2.14}_{-2.14}$\\
		$g_{\Lambda(1890)}$&$-2.83^{+0.12}_{-0.12}$&
		$-2.91^{+0.24}_{-0.24}$&$-3.02^{+0.24}_{-0.24}$\\
		$\chi^2/d.o.f.$&$0.77$&$0.76$&$0.74$\\
		\toprule[0.5pt]
\hline\hline
		\toprule[0.5pt]		
	\end{tabular}	
\end{table}
%%%%%%%%%%%%%%%%%%%%%%
\section{NUMERICAL RESULTS AND DISCUSSION}  \label{sec:MA}
Here, we first determine relevant coupling constants by fitting the $M_{K^{0}_{S} \bar{\Lambda}}$ and $M_{n K^{0}_{S}}$ invariant mass spectra of the $\chi_{c0} \rightarrow n K^{0}_{S} \bar{\Lambda}$ decay. Our estimations find that the invariant mass distributions resulting from below-threshold intermediate states exhibit similar line shapes with no observable structures. In this case, the contributions from the below-threshold states are difficult to  distinguish, and the relevant coupling constants are hardly determined. Therefore, we only include the ground states $N(938)$ and $\Lambda(1116)$ in the present fit. As for the above threshold states, by comparing the structures of the $M_{K^{0}_{S} \bar{\Lambda}}$ and $M_{n K^{0}_{S}}$ invariant mass spectra from the BESIII measurement~\cite{BESIII:2021haw}, we take $N(1720)$, $N(1895)$, $N(2040)$, $\Lambda(1600)$, $\Lambda(1800)$, and $\Lambda(1890)$ as intermediate states, whose $J^{P}$ quantum numbers are $1/2^{\pm}$ or $3/2^{\pm}$. The resonance parameters of the involved states in the present fit are collected in Table~\ref{Tab.1}. Therefore, the total amplitude for the $\chi_{c0} \rightarrow n K^{0}_{S} \bar{\Lambda}$ decay reads,
\begin{eqnarray}
\mathcal{M}_{\mathrm{Tot}}&=&\mathcal{M}_{N(938)}+\mathcal{M}_{N(1720)}+\mathcal{M}_{N(1895)} + \mathcal{M}_{N(2040)} \nonumber\\ &+&\mathcal{M}_{\Lambda(1116)}+\mathcal{M}_{\Lambda(1600)} +\mathcal{M}_{\Lambda(1800)}+\mathcal{M}_{\Lambda(1890)}. \quad
\label{Eq.7}
\end{eqnarray}

With the above total amplitude, one can obtain the differential width of the $\chi_{c0} \rightarrow n K^{0}_{S} \bar{\Lambda}$ decay, which is,
\begin{eqnarray}
	d^{2} \Gamma=\frac{1}{(2J+1)}\frac{1}{(2\pi)^{3}}\frac{1}{32m^{3}}{\bigg|\overline{\mathcal{M}_{\mathrm{Tot}}}\bigg|}^2 dM^{2}_{K^{0}_{S} \bar{\Lambda}} dM^{2}_{n K^{0}_{S}}, \label{Eq.8}
\end{eqnarray}
where $J$ and $m$ are the angular momentum and mass of the initial $\chi_{c0}$ state, respectively. 

As indicated in Eq.~\eqref{Eq.2} and Eq.~\eqref{Eq.7}, the coupling constants $f_{\Lambda N_{J^{P}} K^{0}_{S}}$ and $f_{n N_{J^{P}} \chi_{c0}}$, $f_{n \Lambda_{J^{P}} K^{0}_{S}}$ and $f_{\Lambda \Lambda_{J^{P}} \chi_{c0}}$ always appear in the same amplitude, and the amplitude is proportional to the product of the two coupling constants. Thus, the product of these two coupling constants can be defined as one single parameter, which is,
\begin{eqnarray}
g_{Y}=f_{\Lambda N K^{0}_{S}} \times f_{n N \chi_{c0}}\; (\mathrm{or}\;f_{n \Lambda K^{0}_{S}} \times f_{\Lambda \Lambda \chi_{c0}}) , \label{Eq.11}
\end{eqnarray}
where $Y$ refers to the $N/\Lambda$ baryons and their resonances. The values of $g_{Y}$ will be determined by simultaneously reproducing the $M_{K^{0}_{S} \bar{\Lambda}}$ and $M_{n K^{0}_{S}}$ invariant mass spectra of the $\chi_{c0} \rightarrow n K^{0}_{S} \bar{\Lambda}$ decay.
%%%%%%%%%%%%%%%%%%%%%%
%%%%%%%%%%%%%%%%%%%%%%
\begin{figure*}[htb]
     \includegraphics[width=175mm]{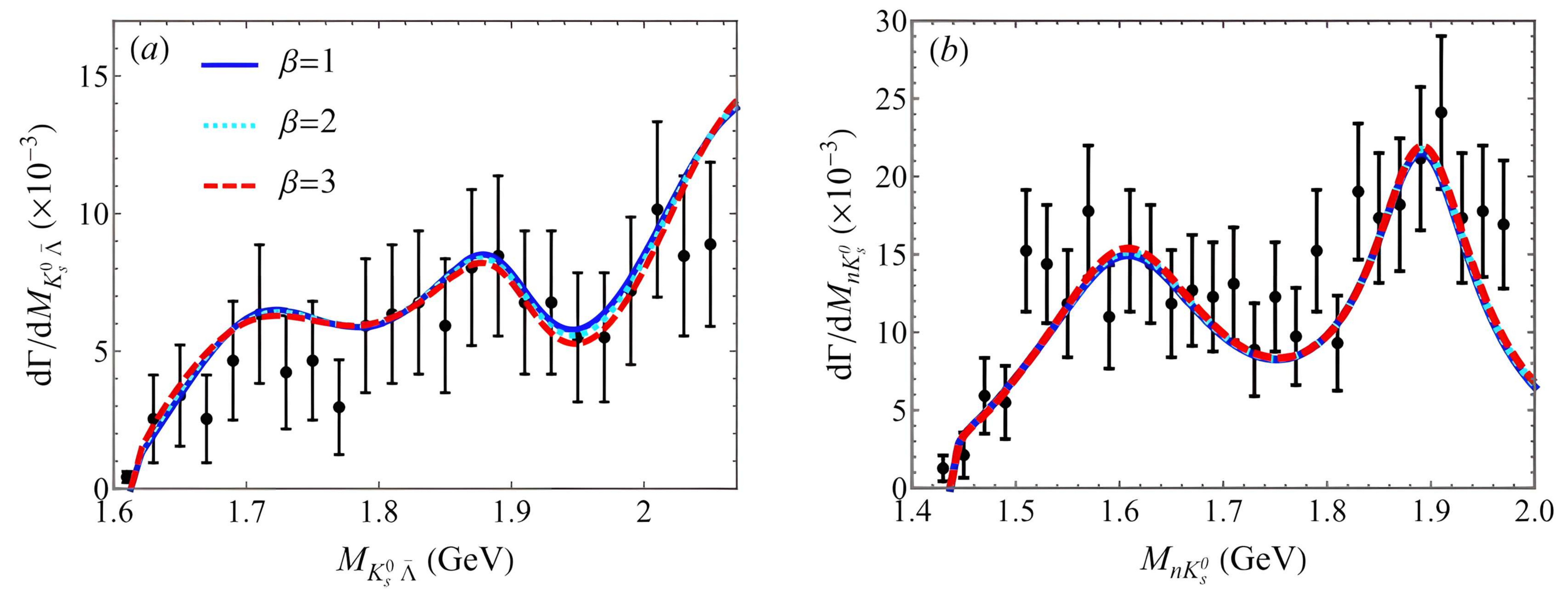}
     \caption{(Color online) The $M_{K^{0}_{S} \bar{\Lambda}}$ (diagram (a)) and $M_{n K^{0}_{S}}$ (diagram (b)) invariant mass spectra of the $\chi_{c0} \rightarrow n K^{0}_{S} \bar{\Lambda}$ decay. The black points with error bars correspond to the experimental data of the $M_{K^{0}_{S} \bar{\Lambda}}$ and $M_{n K^{0}_{S}}$ invariant mass spectra for $\chi_{c0} \rightarrow n K^{0}_{S} \bar{\Lambda}$ from Ref.~\cite{BESIII:2021haw}, while the blue solid, cyan dotted, and red dashed curves are obtained with the fitted parameters for $\beta=1,2,$ and $3$ in Table~\ref{Tab.2}, respectively.}
     \label{Fig.3}
\end{figure*}
%%%%%%%%%%%%%%%%%%%%
%%%%%%%%%%%%%%%%%%%%%%
\begin{figure*}[htb]
     \includegraphics[width=175mm]{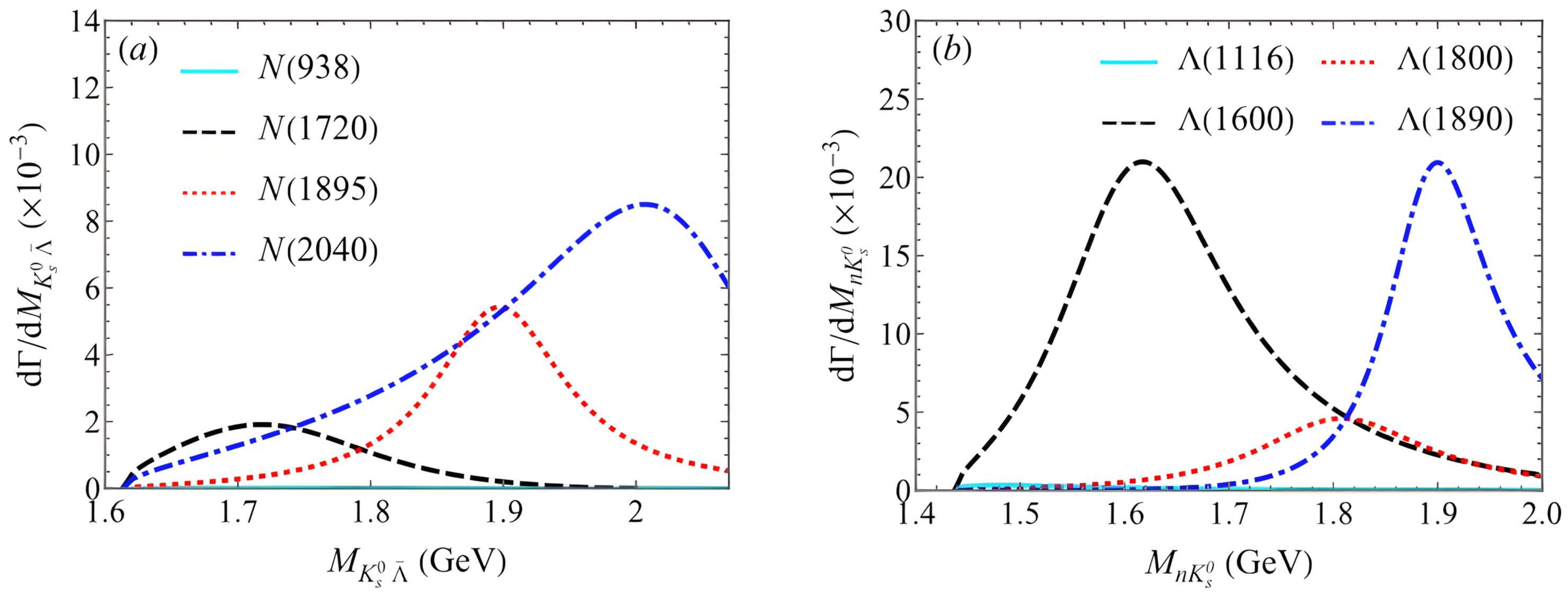}
     \caption{(Color online) The individual contributions from different intermediate states for the $M_{K^{0}_{S} \bar{\Lambda}}$ (diagram (a)) and $M_{n K^{0}_{S}}$ (diagram (b)) invariant mass spectra for $\chi_{c0} \rightarrow n K^{0}_{S} \bar{\Lambda}$, where we take the parameter $\beta=1$.}
     \label{Fig.4}
\end{figure*}

%%%%%%%%%%%%%%%%%%%%%%
\begin{figure*}[htb]
     \includegraphics[width=175mm]{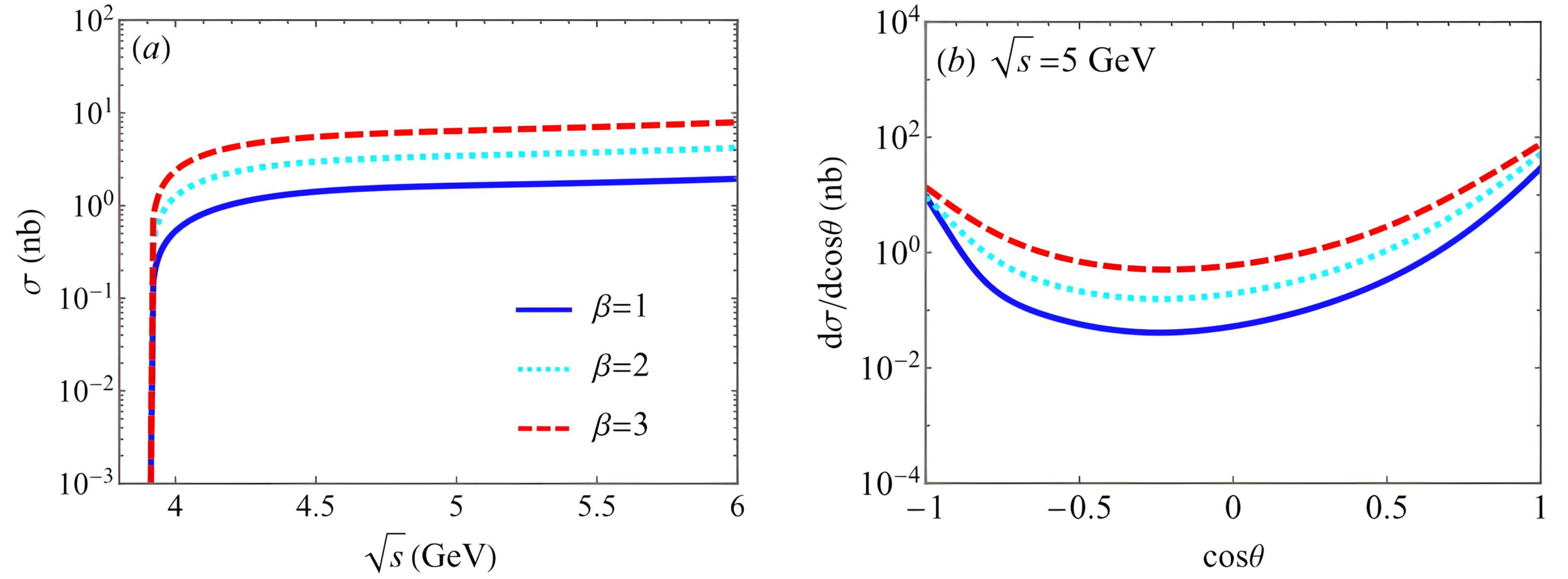}
     \caption{(Color online) Diagram (a) corresponds to the cross sections for the $\bar{\Lambda} n \rightarrow \chi_{c0} K^{0}_{S}$ process depending on the $\sqrt{s}$. Diagram (b) corresponds to the differential cross sections depending on cos$\theta$ with $\sqrt{s}=5$ GeV. The blue solid, cyan dotted, and red dashed curves are obtained with the fitted parameters for $\beta=1,2,$ and $3$ in Table~\ref{Tab.2}, respectively.} 
     \label{Fig.5}
\end{figure*}
%%%%%%%%%%%%%%%%%%%%

%%%%%%%%%%%%%%%%%%%%

\subsection{The $\chi_{c0} \rightarrow n K^{0}_{S} \bar{\Lambda}$ decay}
In the present work, the parameter $g_Y$  can be determined by fitting the $M_{K^{0}_{S} \bar{\Lambda}}$ and $M_{n K^{0}_{S}}$ invariant mass spectra of the $\chi_{c0} \rightarrow n K^{0}_{S} \bar{\Lambda}$ decay. With typical $\beta$ values, we list the fitted parameter values in Table~\ref{Tab.2}. With the central values of the fitted parameters, one has $\chi^2 /d.o.f.$ $=$ 0.77, 0.76, and 0.74 for $\beta=1,\ 2,$ and $3$, respectively, which are acceptable for the present fit. Notably, all the intermediate states are included simultaneously in the present fit, and the relevant $g_Y$ are all considered as free parameters. Theoretically, for the two ground states, $N(938)$ and $\Lambda(1116)$, the coupling constants $f_{n N \chi_{c0}}$ and $f_{\Lambda \Lambda \chi_{c0}}$ can be determined by combining the related effective Lagrangians with the decay widths of the $\chi_{c0} \to p \bar{p}$ and $\chi_{c0} \to \Lambda \bar{\Lambda}$ processes~\cite{ParticleDataGroup:2026mpi}, respectively, while the coupling constant $f_{\Lambda N K^{0}_{S}}$ can be determined by SU(3) symmetry~\cite{Ronchen:2012eg}. However, in addition to the ground states, there also exist some excited $N^{\ast}$ and $\Lambda^{\ast}$ states below the $K_S^0 \bar{\Lambda}$ and $nK_S^0$ thresholds, whose line shapes in the invariant mass spectra are nearly identical to those of the $N(938)$ and $\Lambda(1116)$. Therefore, the parameter values estimated from $\chi_{c0}$ decay and SU(3) symmetry should be different with the present fit.

In Fig.~\ref{Fig.3}, we present the fitted results for $M_{K^{0}_{S} \bar{\Lambda}}$ [Fig.~\ref{Fig.3}(a)] and $M_{n K^{0}_{S}}$ [Fig.~\ref{Fig.3}(b)] invariant mass spectra of the $\chi_{c0} \rightarrow n K^{0}_{S} \bar{\Lambda}$ decay, where the black points with error bars correspond to the experimental data from Ref.~\cite{BESIII:2021haw}. The blue solid, cyan dotted, and red dashed curves are obtained with the fitted parameters for $\beta=1,2,$ and $3$ in Table~\ref{Tab.2}, respectively. From the figure one can find that the fitted $M_{K^{0}_{S} \bar{\Lambda}}$ and $M_{n K^{0}_{S}}$ invariant mass spectra reproduce the experimental data well over the entire considered range. Thus, although the parameters $g_{Y}$ vary with $\beta$, the overall fit results remain consistent. Moreover, our results show that there are two clear structures near 1.70 $\mathrm{GeV}$ and 1.90 $\mathrm{GeV}$ in the $M_{K^{0}_{S} \bar{\Lambda}}$ invariant mass spectrum,
while two additional structures are observed near 1.60 $\mathrm{GeV}$ and 1.90 $\mathrm{GeV}$ in the $M_{n K^{0}_{S}}$ invariant mass spectrum.

In addition to the total invariant mass spectra, the individual contributions of each intermediate state for $\chi_{c0} \rightarrow n K^{0}_{S} \bar{\Lambda}$ are presented in Fig.~\ref{Fig.4}, where the cyan solid, black dashed, red dotted, and blue dash-dotted curves correspond to the contributions from $N(938)/\Lambda(1116)$, $N(1720)/\Lambda(1600)$, $N(1895)/\Lambda(1800)$, and $N(2040)/\Lambda(1890)$, respectively. In Fig.~\ref{Fig.4}-(a), one can find that the contribution from the $N(938)$ intermediate state is much smaller than those from the other three nucleon resonances. Moreover, the signals of $N(1720)$ and $N(1895)$ can be observed clearly, which indicate that the two structures near 1.70 $\mathrm{GeV}$ and 1.90 $\mathrm{GeV}$ in the $M_{K^{0}_{S} \bar{\Lambda}}$ invariant mass spectrum should correspond to $N(1720)$ and $N(1895)$, respectively. In Fig.~\ref{Fig.4}-(b), we present the individual contributions from $\Lambda(1116)$, $\Lambda(1600)$, $\Lambda(1800)$, and $\Lambda(1890)$. Similar to the case of $N(938)$, the contribution from $\Lambda(1116)$ is relatively small. Moreover, the signals of $\Lambda(1600)$ and $\Lambda(1890)$ are both clearly visible, which are several times larger than those of $\Lambda(1800)$. This indicates that the structures near 1.60 $\mathrm{GeV}$ and 1.90 $\mathrm{GeV}$ in the $M_{n K^{0}_{S}}$ invariant mass spectrum should correspond to $\Lambda(1600)$ and $\Lambda(1890)$.

%%%%%%%%%%%%%%%%%%%%%%
\begin{figure*}[htb]
     \includegraphics[width=175mm]{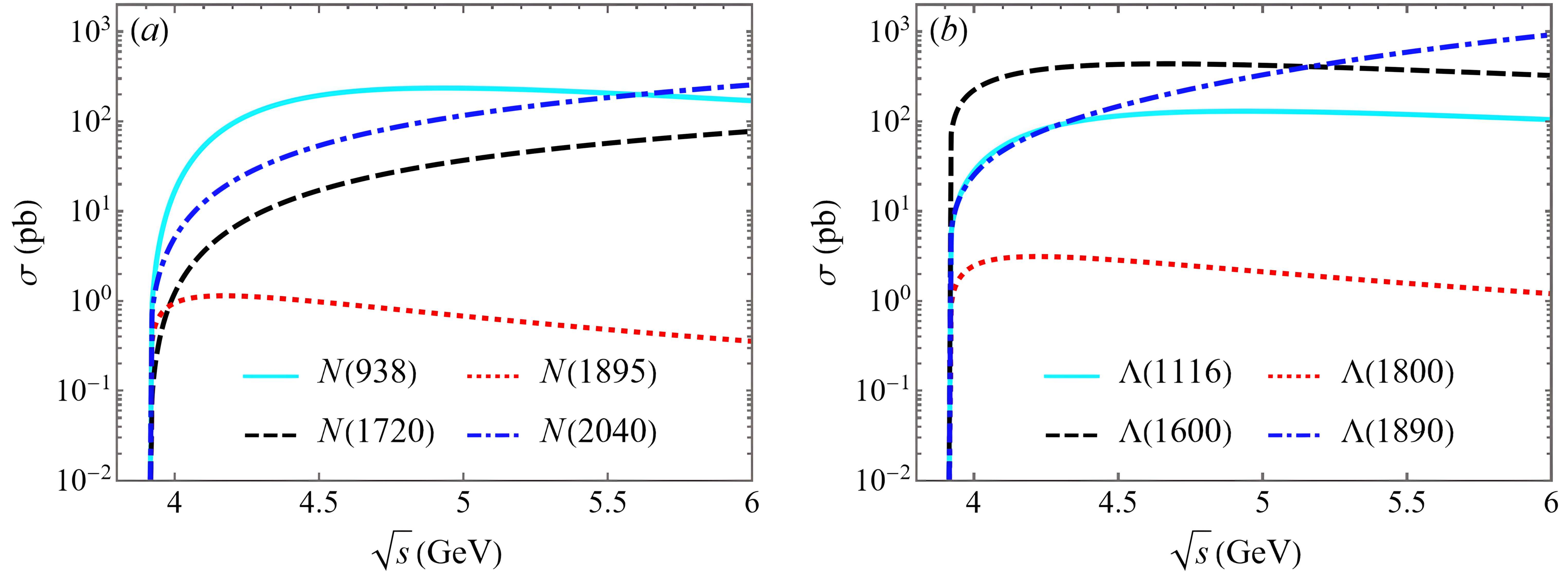}
     \caption{(Color online) The individual contributions from different intermediate states for the cross sections for the $\bar{\Lambda} n \rightarrow \chi_{c0} K^{0}_{S}$ process depending on the $\sqrt{s}$, where the parameter $\beta$ is set to be 1.}
     \label{Fig.6}
\end{figure*}
%%%%%%%%%%%%%%%%%%%%

\subsection{Cross sections for $\bar{\Lambda} n \rightarrow \chi_{c0} K^{0}_{S}$}
After determining the parameters $g_{Y}$, we can further estimate the cross sections for the $\bar{\Lambda} n \rightarrow \chi_{c0} K^{0}_{S}$ process. With the amplitudes in Eq.~\eqref{Eq.6}, the total amplitude of $\bar{\Lambda} n \rightarrow \chi_{c0} K^{0}_{S}$ can be expressed as,
\begin{eqnarray}
\mathcal{M}^{\prime}_{\mathrm{Tot}}&=&\mathcal{M}^{t}_{N(938)}+\mathcal{M}^{t}_{N(1720)}+\mathcal{M}^{t}_{N(1895)} + \mathcal{M}^{t}_{N(2040)} \nonumber\\ &+&\mathcal{M}^{u}_{\Lambda(1116)}+\mathcal{M}^{u}_{\Lambda(1600)} +\mathcal{M}^{u}_{\Lambda(1800)}+\mathcal{M}^{u}_{\Lambda(1890)}. \quad \label{Eq.a}
\end{eqnarray}
With the above total amplitude, the differential cross sections depending on $\cos\theta$ for $\bar{\Lambda} n \rightarrow \chi_{c0} K^{0}_{S}$ can be obtained by
\begin{eqnarray}
	\frac{d{\sigma}} {d \cos\theta}
	=\frac{1} {32 \pi s} \frac{|\vec{p}_f|} {|\vec{p}_i|}  \left(\frac{1} {4} {\bigg|\overline{\mathcal{M}^{\prime}_{\mathrm{Tot}}}\bigg|}^2\right), \label{Eq.b}
\end{eqnarray}
where $s$ refers to the square of the center-of-mass energy. $\theta$ is the scattering angle, which is the angle of outgoing $K^{0}_{S}$ and the incident $\bar{\Lambda}$ beam direction in the center-of-mass frame. The $\vec{p_{f}}$ and $\vec{p_{i}}$ are the three-momentum of the final $K^{0}_{S}$ meson and the initial $\bar{\Lambda}$ beam in the center-of-mass system, respectively.

With the above preparations, the cross sections for $\bar{\Lambda} n \rightarrow \chi_{c0} K^{0}_{S}$ depending on the $\sqrt{s}$ are presented in Fig.~\ref{Fig.5}-(a), where the blue solid, cyan dotted, and red dashed curves are obtained with the fitted parameters for $\beta=1,2,$ and $3$ in Table~\ref{Tab.2}, respectively. From the figure, one can find that each cross section curve increases rapidly near the threshold and then becomes weakly dependent on the $\sqrt{s}$. At $\sqrt{s}=6$ GeV, the cross sections are estimated to be 1.9 nb, 4.2 nb, and 8.0 nb when $\beta=1,2,$ and $3$, respectively. In addition to the cross sections, the differential cross sections with $\sqrt{s}$ = 5 GeV for $\bar{\Lambda} n \rightarrow \chi_{c0} K^{0}_{S}$ are presented in Fig.~\ref{Fig.5}-(b). Similar to  Fig.~\ref{Fig.5}-(a), we present the results obtained with the fitted parameters for $\beta=1$, $2$, and $3$ in Table~\ref{Tab.2}. From the figure, our estimations show that the differential cross sections are much larger in the forward and backward angle regions, and reach the maximum at the forward angle region. In addition, the differential cross sections become more concentrated in the forward and backward regions as $\beta$ increases.

Besides the total cross sections, the individual contributions of each intermediate state for $\bar{\Lambda} n \rightarrow \chi_{c0} K^{0}_{S}$ are presented in Fig.~\ref{Fig.6}, where the cyan solid, black dashed, red dotted, and blue dash-dotted curves correspond to the contributions from $N(938)/\Lambda(1116)$, $N(1720)/\Lambda(1600)$, $N(1895)/\Lambda(1800)$, and $N(2040)/\Lambda(1890)$, respectively. Here we take the parameter $\beta=1$ as an example. From this figure, one can find that the cross sections resulting from $N(1895)$ and $\Lambda(1800)$ are about two orders of magnitude smaller than those from the other intermediate states, which indicates that their contributions are negligible. In addition, the cross sections resulted from $N(938)$, $\Lambda(1116)$, $\Lambda(1600)$, $N(1720)$, and $N(2040)$ are estimated to be  $\mathcal{O}$($10^{2}$) pb at $\sqrt{s}=6$ GeV, while the contributions from the two ground states account for about 15$\%$ of the total cross section. Meanwhile, our estimations indicate that the cross section of $\Lambda(1890)$ is about 1 nb at $\sqrt{s}=6$ GeV, accounting for about half of the total cross section.

\section{SUMMARY}
In recent years, (anti-)hyperon-nucleon scattering has been actively studied, particularly with the advancement and upgrading of experimental facilities. Motivated by the $\chi_{c0} \rightarrow n K^{0}_{S} \bar{\Lambda}$ decay reported from BESIII Collaboration, we propose to study the $\bar{\Lambda} n \rightarrow \chi_{c0} K^{0}_{S}$ process using an effective Lagrangian approach.

In the present work, the coupling constants are determined by fitting the $M_{K^{0}_{S} \bar{\Lambda}}$ and $M_{n K^{0}_{S}}$ invariant mass spectra of the $\chi_{c0} \rightarrow n K^{0}_{S} \bar{\Lambda}$ decay, where eight $N$ and $\Lambda$ resonances are considered. Our estimations indicate that the contributions from $N(938)$ and $\Lambda(1116)$ are relatively small. In addition, two clear structures near 1.70~GeV and 1.90~GeV are observed in the $M_{K^{0}_{S} \bar{\Lambda}}$ spectrum, which correspond to $N(1720)$ and $N(1895)$, respectively. Moreover, the structures near 1.60 GeV and 1.90 GeV in the $M_{n K^{0}_{S}}$ invariant mass spectrum correspond to $\Lambda(1600)$ and $\Lambda(1890)$, respectively.

Using the fitted parameters $g_{Y}$, we investigate the production of $\chi_{c0}$ in the $\bar{\Lambda} n \rightarrow \chi_{c0} K^{0}_{S}$ process. Our estimations show that the cross sections increase rapidly near the threshold and then become weakly dependent on the $\sqrt{s}$. At $\sqrt{s}=6$ GeV, the cross sections for $\bar{\Lambda} n \rightarrow \chi_{c0} K^{0}_{S}$ are estimated to be 1.9 nb, 4.2 nb, and 8.0 nb for $\beta=1$, 2, and 3, respectively. For the individual contributions, our results show that the contributions from the two ground states $N(938)$ and $\Lambda(1116)$ account for about 15$\%$ of the total cross section. Moreover, our estimations indicate that the cross section resulting from $\Lambda(1890)$ plays an important role, accounting for about half of the total cross section. Furthermore, we estimate the differential cross sections for $\bar{\Lambda} n \rightarrow \chi_{c0} K^{0}_{S}$ at $\sqrt{s}=5$ GeV, which could be tested by future experiments.

\section*{ACKNOWLEDGMENTS}
This work is partly supported by the National Natural Science Foundation of China under the Grant Nos. 12175037 and 12335001, as well as supported, in part, by National Key Research and Development Program under the contract No. 2024YFA1610503. Quan-Yun Guo is also supported by the SEU Innovation Capability Enhancement Plan for Doctoral Students (Grant No. CXJH SEU 26160).

%%%%%%%%%
 
\bibliographystyle{apsrev4-1}
\bibliography{LambdaBarN.bib}

@article{Machleidt:1987hj,
    author = "Machleidt, R. and Holinde, K. and Elster, C.",
    title = "{The Bonn Meson Exchange Model for the Nucleon Nucleon Interaction}",
    journal = "Phys. Rept.",
    volume = "149",
    pages = "1--89",
    year = "1987",
    doi = "10.1016/S0370-1573(87)80002-9"
}

@article{Kaiser:1997mw,
    author = "Kaiser, N. and Brockmann, R. and Weise, W.",
    title = "{Peripheral nucleon-nucleon phase shifts and chiral symmetry}",
    journal = "Nucl. Phys. A",
    volume = "625",
    pages = "758--788",
    year = "1997",
    doi = "10.1016/S0375-9474(97)00586-1",
    archivePrefix = "arXiv",
    eprint = "nucl-th/9706045",
    primaryClass = "nucl-th"
}

@article{Kaplan:1996xu,
    author = "Kaplan, D. B. and Savage, M. J. and Wise, M. B.",
    title = "{Nucleon-nucleon scattering from effective field theory}",
    journal = "Nucl. Phys. B",
    volume = "478",
    pages = "629--659",
    year = "1996",
    doi = "10.1016/0550-3213(96)00357-4",
    archivePrefix = "arXiv",
    eprint = "nucl-th/9605002",
    primaryClass = "nucl-th"
}

@article{Nakayama:1988zz,
    author = "Nakayama, K. and Love, W. G.",
    title = "{Effective interaction for nucleon-nucleus scattering based on a one-boson-exchange model}",
    journal = "Phys. Rev. C",
    volume = "38",
    pages = "51--73",
    year = "1988",
    doi = "10.1103/PhysRevC.38.51"
}

@article{Williams:1987yn,
    author = "Williams, A. G. and Thomas, A. W. and Miller, G. A.",
    title = "{Charge Symmetry Breaking in Neutron-Proton Elastic Scattering}",
    journal = "Phys. Rev. C",
    volume = "36",
    pages = "1956",
    year = "1987",
    doi = "10.1103/PhysRevC.36.1956"
}

@article{Alexander:1966ara,
    author = "Alexander, G. and Benary, O. and Karshon, U. and Shapira, A. and Yukutieli, G. and Engelmann, R. and Filthuth, H. and Fridman, A. and Schiby, B.",
    title = "{Low energy $\Lambda p \to \Lambda p$ cross section measurement and determination of S-wave scattering lengths}",
    journal = "Phys. Lett.",
    volume = "19",
    pages = "715--719",
    year = "1966",
    doi = "10.1016/0031-9163(66)90456-2"
}

@article{Alexander:1968acu,
    author = "Alexander, G. and Karshon, U. and Shapira, A. and Yekutieli, G. and Engelmann, R. and Filthuth, H. and Lughofer, W.",
    title = "{Study of the $\Lambda-N$ System in Low-Energy $\Lambda-p$ Elastic Scattering}",
    journal = "Phys. Rev.",
    volume = "173",
    pages = "1452--1460",
    year = "1968",
    doi = "10.1103/PhysRev.173.1452"
}

@article{Sechi-Zorn:1968mao,
    author = "Sechi-Zorn, B. and Kehoe, B. and Twitty, J. and Burnstein, R. A.",
    title = "{Low-energy $\Lambda$-Proton Elastic Scattering}",
    journal = "Phys. Rev.",
    volume = "175",
    pages = "1735--1740",
    year = "1968",
    doi = "10.1103/PhysRev.175.1735"
}

@article{Bassano:1967kbh,
    author = "Bassano, D. and Chang, C. Y. and Goldberg, M. and Kikuchi, T. and Leitner, J.",
    title = "{Lambda-Proton Interactions at High Energies}",
    journal = "Phys. Rev.",
    volume = "160",
    number = "5",
    pages = "1239",
    year = "1967",
    doi = "10.1103/PhysRev.160.1239"
}

@article{Alexander:1969tv,
    author = "Alexander, G. and Benary, O. and Karshon, U. and Maor, U.",
    title = "{Data compilation of baryon baryon interactions (iii) - hyperon-proton collisions}",
    journal = "Nucl. Phys. B",
    volume = "10",
    pages = "554--562",
    year = "1969",
    doi = "10.1016/0550-3213(69)90041-8"
}

@article{Charlton:1970bv,
    author = "Charlton, G. R. and Badier, J. and Barrelet, E. and Makarovisch, I. and Pernegr, J. and Hubbard, J. R. and Leveque, A. and Louedec, C. and Moscoso, L. and Revel, D.",
    title = "{Hyperon-proton interactions between 0.5 and 4.0 GeV/c}",
    journal = "Phys. Lett. B",
    volume = "32",
    pages = "720--722",
    year = "1970",
    doi = "10.1016/0370-2693(70)90454-5"
}

@article{Hauptman:1977hr,
    author = "Hauptman, J. M. and Kadyk, J. A. and Trilling, G. H.",
    title = "{Experimental Study of $\Lambda p$ and $\Xi^{0} p$ Interactions in the Range 1-10 GeV/c}",
    journal = "Nucl. Phys. B",
    volume = "125",
    pages = "29--51",
    year = "1977",
    doi = "10.1016/0550-3213(77)90222-X"
}

@article{Kubis:1970im,
    author = "Kubis, J. J. and Walters, H. R. J.",
    title = "{Regge-pole model for $\Lambda n$ interactions}",
    journal = "Nucl. Phys. B",
    volume = "17",
    pages = "547--564",
    year = "1970",
    doi = "10.1016/0550-3213(70)90011-8"
}

@article{Anderson:1975rh,
    author = "Anderson, K. J. and Gelfand, N. M. and Keren, J. and Tan, T. H.",
    title = "{$\Lambda$-proton elastic scattering from 1 to 17 GeV/c}",
    journal = "Phys. Rev. D",
    volume = "11",
    pages = "473--477",
    year = "1975",
    doi = "10.1103/PhysRevD.11.473",
    note = "[erratum: Phys. Rev. D 11, 3359 (1975)]"
}

@article{Mount:1975pb,
    author = "Mount, R. P. and Ansorge, R. E. and Carter, J. R. and Neale, W. W. and Rushbrooke, J. G. and Ward, D. R.",
    title = "{$\Lambda p$ Interactions Below 24 GeV/c}",
    journal = "Phys. Lett. B",
    volume = "58",
    pages = "228--232",
    year = "1975",
    doi = "10.1016/0370-2693(75)90642-5"
}

@article{Giacomelli:1976gn,
    author = "Giacomelli, G.",
    title = "{Total Cross-Sections and Elastic Scattering at High-Energies.}",
    journal = "Phys. Rept.",
    volume = "23",
    pages = "123--235",
    year = "1976",
    doi = "10.1016/0370-1573(76)90030-2"
}

@article{CLAS:2021gur,
    author = "Rowley, J. and others",
    collaboration = "CLAS",
    title = "{Improved $\Lambda p$ Elastic Scattering Cross Sections Between 0.9 and 2.0 GeV/c and Connections to the Neutron Star Equation of State}",
    journal = "Phys. Rev. Lett.",
    volume = "127",
    number = "27",
    pages = "272303",
    year = "2021",
    archivePrefix = "arXiv",
    eprint = "2108.03134",
    primaryClass = "hep-ex"
}

@article{BESIII:2024geh,
    author = "Ablikim, M. and others",
    collaboration = "BESIII",
    title = "{First study of antihyperon-nucleon scattering $\bar{\Lambda}p \to \bar{\Lambda}p$ and Measurement of $\Lambda p \to \Lambda p$ Cross Section}",
    journal = "Phys. Rev. Lett.",
    volume = "132",
    number = "23",
    pages = "231902",
    year = "2024",
    archivePrefix = "arXiv",
    eprint = "2401.09012",
    primaryClass = "hep-ex"
}

@misc{BESIII:2025bft,
    author = "Ablikim, M. and others",
    collaboration = "BESIII",
    title = "{First measurement of $\Sigma^{+}n \rightarrow \Lambda p$ and $\Sigma^{+}n \rightarrow \Sigma^{0} p$ cross sections via $\Sigma^{+}$-nucleus scattering at an electron-positron collider}",
    year = "2025",
    archivePrefix = "arXiv",
    eprint = "2505.19907",
    primaryClass = "hep-ex"
}

@misc{BESIII:2025vqk,
    author = "Ablikim, M. and others",
    collaboration = "BESIII",
    title = "{Study of the reaction $\Xi^{0}n \rightarrow \Lambda \Lambda X$ using $\Xi^{0}$-nucleus scattering}",
    year = "2025",
    archivePrefix = "arXiv",
    eprint = "2512.04701",
    primaryClass = "hep-ex"
}

@article{Holzenkamp:1989tq,
    author = "Holzenkamp, B. and Holinde, K. and Speth, J.",
    title = "{A Meson Exchange Model for the Hyperon-Nucleon Interaction}",
    journal = "Nucl. Phys. A",
    volume = "500",
    pages = "485--528",
    year = "1989",
    doi = "10.1016/0375-9474(89)90223-6"
}

@article{Rijken:1998yy,
    author = "Rijken, T. A. and Stoks, V. G. J. and Yamamoto, Y.",
    title = "{Soft core hyperon-nucleon potentials}",
    journal = "Phys. Rev. C",
    volume = "59",
    pages = "21--40",
    year = "1999",
    doi = "10.1103/PhysRevC.59.21",
    archivePrefix = "arXiv",
    eprint = "nucl-th/9807082",
    primaryClass = "nucl-th"
}

@article{Tominaga:2001ra,
    author = "Tominaga, K. and Ueda, T.",
    title = "{Effective one-boson-exchange potential for $\Lambda N$ and $\Sigma N$ systems and hypertriton}",
    journal = "Nucl. Phys. A",
    volume = "693",
    pages = "731--754",
    year = "2001",
    doi = "10.1016/S0375-9474(01)00882-X"
}

@article{Haidenbauer:2005zh,
    author = "Haidenbauer, J. and Meissner, U. G.",
    title = "{The Julich hyperon-nucleon model revisited}",
    journal = "Phys. Rev. C",
    volume = "72",
    pages = "044005",
    year = "2005",
    doi = "10.1103/PhysRevC.72.044005",
    archivePrefix = "arXiv",
    eprint = "nucl-th/0506019",
    primaryClass = "nucl-th"
}

@article{Polinder:2006zh,
    author = "Polinder, H. and Haidenbauer, J. and Meissner, U. G.",
    title = "{Hyperon-nucleon interactions: a Chiral effective field theory approach}",
    journal = "Nucl. Phys. A",
    volume = "779",
    pages = "244--266",
    year = "2006",
    doi = "10.1016/j.nuclphysa.2006.09.006",
    archivePrefix = "arXiv",
    eprint = "nucl-th/0605050",
    primaryClass = "nucl-th"
}

@article{Haidenbauer:2013oca,
    author = "Haidenbauer, J. and Petschauer, S. and Kaiser, N. and Meissner, U. G. and Nogga, A. and Weise, W.",
    title = "{Hyperon-nucleon interaction at next-to-leading order in chiral effective field theory}",
    journal = "Nucl. Phys. A",
    volume = "915",
    pages = "24--58",
    year = "2013",
    doi = "10.1016/j.nuclphysa.2013.06.008",
    archivePrefix = "arXiv",
    eprint = "1304.5339",
    primaryClass = "nucl-th"
}

@article{Li:2016mln,
    author = "Li, K. W. and Ren, X. L. and Geng, L. S. and Long, B. W.",
    title = "{Leading order relativistic hyperon-nucleon interactions in chiral effective field theory}",
    journal = "Chin. Phys. C",
    volume = "42",
    number = "1",
    pages = "014105",
    year = "2018",
    doi = "10.1088/1674-1137/42/1/014105",
    archivePrefix = "arXiv",
    eprint = "1612.08482",
    primaryClass = "nucl-th"
}

@article{Haidenbauer:2019boi,
    author = "Haidenbauer, J. and Mei{\ss}ner, U. G. and Nogga, A.",
    title = "{Hyperon–nucleon interaction within chiral effective field theory revisited}",
    journal = "Eur. Phys. J. A",
    volume = "56",
    number = "3",
    pages = "91",
    year = "2020",
    doi = "10.1140/epja/s10050-020-00100-4",
    archivePrefix = "arXiv",
    eprint = "1906.11681",
    primaryClass = "nucl-th"
}

@article{Song:2021yab,
    author = "Song, J. and Liu, Z. W. and Li, K. W. and Geng, L. S.",
    title = "{Test of the hyperon-nucleon interaction within leading order covariant chiral effective field theory}",
    journal = "Phys. Rev. C",
    volume = "105",
    number = "3",
    pages = "035203",
    year = "2022",
    doi = "10.1103/PhysRevC.105.035203",
    archivePrefix = "arXiv",
    eprint = "2107.04742",
    primaryClass = "nucl-th"
}

@article{Haidenbauer:2023qhf,
    author = "Haidenbauer, J. and Mei{\ss}ner, U. G. and Nogga, A. and Le, H.",
    title = "{Hyperon–nucleon interaction in chiral effective field theory at next-to-next-to-leading order}",
    journal = "Eur. Phys. J. A",
    volume = "59",
    number = "3",
    pages = "63",
    year = "2023",
    doi = "10.1140/epja/s10050-023-00960-6",
    archivePrefix = "arXiv",
    eprint = "2301.00722",
    primaryClass = "nucl-th"
}

@article{Wang:2024whi,
    author = "Wang, X. Y. and Gao, Y. and Liu, X.",
    title = "{Understanding of the BESIII measurement of (anti)hyperon-nucleon scattering}",
    journal = "Phys. Lett. B",
    volume = "862",
    pages = "139321",
    year = "2025",
    archivePrefix = "arXiv",
    eprint = "2410.17842",
    primaryClass = "hep-ph"
}

@article{Chen:2025qxh,
    author = "Chen, Y. and He, J.",
    title = "{Study of $\Lambda p$ and $\bar{\Lambda} p$ scatterings via quasipotential Bethe-Salpeter equation}",
    journal = "Phys. Lett. B",
    volume = "871",
    pages = "140004",
    year = "2025",
    doi = "10.1016/j.physletb.2025.140004",
    archivePrefix = "arXiv",
    eprint = "2507.18415",
    primaryClass = "hep-ph"
}

@misc{Liu:2026gxr,
    author = "Liu, H. and Liu, L. and Tan, J. X. and Wang, W. and Yan, H. and Zhu, Q. T.",
    title = "{A Lattice QCD study of $p-\Lambda$ scattering in continuum and chiral limits}",
    year = "2026",
    archivePrefix = "arXiv",
    eprint = "2603.05854",
    primaryClass = "hep-lat"
}

@article{Lundborg:2005am,
    author = "Lundborg, A. and Barnes, T. and Wiedner, U.",
    title = "{Charmonium production in p anti-p annihilation: Estimating cross sections from decay widths}",
    journal = "Phys. Rev. D",
    volume = "73",
    pages = "096003",
    year = "2006",
    doi = "10.1103/PhysRevD.73.096003",
    archivePrefix = "arXiv",
    eprint = "hep-ph/0507166",
    primaryClass = "hep-ph"
}

@article{Xu:2015qqa,
    author = "Xu, H. and Xie, J. J. and Liu, X.",
    title = "{Implication of the observed $e^{+}e^{-}\rightarrow p{\bar{p}}\pi ^0$ for studying the $p{\bar{p}}\rightarrow \psi (3770)\pi ^0$ process}",
    journal = "Eur. Phys. J. C",
    volume = "76",
    number = "4",
    pages = "192",
    year = "2016",
    doi = "10.1140/epjc/s10052-016-4054-4",
    archivePrefix = "arXiv",
    eprint = "1505.04571",
    primaryClass = "hep-ph"
}

@article{Wang:2017sxq,
    author = "Wang, J. Z. and Xu, H. and Xie, J. J. and Liu, X.",
    title = "{Production of the charmoniumlike state Y(4220) through the $p\bar{p} \to Y(4220) \pi^0$ reaction}",
    journal = "Phys. Rev. D",
    volume = "96",
    number = "9",
    pages = "094004",
    year = "2017",
    doi = "10.1103/PhysRevD.96.094004",
    archivePrefix = "arXiv",
    eprint = "1710.08738",
    primaryClass = "hep-ph"
}

@article{Zhou:2020smj,
    author = "Zhou, Q. S. and Wang, J. Z. and Xie, J. J. and Liu, X.",
    title = "{Role of the low-lying nucleon resonances in the $p\bar{p} \to \psi \eta$ reaction}",
    journal = "Phys. Rev. D",
    volume = "101",
    number = "11",
    pages = "114001",
    year = "2020",
    doi = "10.1103/PhysRevD.101.114001",
    archivePrefix = "arXiv",
    eprint = "2003.07135",
    primaryClass = "hep-ph"
}

@article{BESIII:2021haw,
    author = "Ablikim, M. and others",
    collaboration = "BESIII",
    title = "{Observation of the decays $\chi_{cJ}\to n K^{0}_{S}\bar\Lambda + c.c.$}",
    journal = "JHEP",
    volume = "11",
    pages = "217",
    year = "2021",
    archivePrefix = "arXiv",
    eprint = "2106.13442",
    primaryClass = "hep-ex"
}

@article{Chen:2011xk,
    author = "Chen, D. Y. and Liu, X.",
    title = "{Predicted charged charmonium-like structures in the hidden-charm dipion decay of higher charmonia}",
    journal = "Phys. Rev. D",
    volume = "84",
    pages = "034032",
    year = "2011",
    archivePrefix = "arXiv",
    eprint = "1106.5290",
    primaryClass = "hep-ph"
}

@article{Okubo:1975sc,
    author = "Okubo, S.",
    title = "{SU(4), SU(8) Mass Formulas and Weak Interactions}",
    journal = "Phys. Rev. D",
    volume = "11",
    pages = "3261--3269",
    year = "1975"
}

@article{Jackson:2015dva,
    author = "Jackson, B. C. and Oh, Y. and Haberzettl, H. and Nakayama, K.",
    title = "{$\bar{K} + N \to K + \Xi$ reaction and $S=-1$ hyperon resonances}",
    journal = "Phys. Rev. C",
    volume = "91",
    number = "6",
    pages = "065208",
    year = "2015",
    archivePrefix = "arXiv",
    eprint = "1503.00845",
    primaryClass = "nucl-th"
}

@article{Oh:2000qr,
    author = "Oh, Y. s. and Song, T. and Lee, S. H.",
    title = "{$J/\psi$ absorption by $\pi$ and $\rho$ mesons in meson exchange model with anomalous parity interactions}",
    journal = "Phys. Rev. C",
    volume = "63",
    pages = "034901",
    year = "2001",
    archivePrefix = "arXiv",
    eprint = "nucl-th/0010064",
    primaryClass = "nucl-th"
}

@article{Tsushima:1996xc,
    author = "Tsushima, K. and Sibirtsev, A. and Thomas, A. W.",
    title = "{Resonance model study of strangeness production in $pp$ collisions}",
    journal = "Phys. Lett. B",
    volume = "390",
    pages = "29--35",
    year = "1997",
    archivePrefix = "arXiv",
    eprint = "nucl-th/9608029",
    primaryClass = "nucl-th"
}

@article{Tsushima:1996tv,
    author = "Tsushima, K. and Huang, S. W. and Faessler, A.",
    title = "{A Study of the $\pi B \to Y K$ reactions for kaon production in heavy ion collisions}",
    journal = "Austral. J. Phys.",
    volume = "50",
    pages = "35--43",
    year = "1997",
    archivePrefix = "arXiv",
    eprint = "nucl-th/9602005",
    primaryClass = "nucl-th"
}

@article{Tsushima:1998jz,
    author = "Tsushima, K. and Sibirtsev, A. and Thomas, A. W. and Li, G. Q.",
    title = "{Resonance model study of kaon production in baryon baryon reactions for heavy ion collisions}",
    journal = "Phys. Rev. C",
    volume = "59",
    pages = "369--387",
    year = "1999",
    note = "[erratum: Phys. Rev. C 61, 029903 (2000)]",
    archivePrefix = "arXiv",
    eprint = "nucl-th/9801063",
    primaryClass = "nucl-th"
}

@article{Ouyang:2009kv,
    author = "Ouyang, Z. and Xie, J. J. and Zou, B. S. and Xu, H. S.",
    title = "{Theoretical study on $pp \to p n \pi^{+}$ reaction at medium energies}",
    journal = "Int. J. Mod. Phys. E",
    volume = "18",
    pages = "281--292",
    year = "2009",
    archivePrefix = "arXiv",
    eprint = "0902.1818",
    primaryClass = "nucl-th"
}

@article{ParticleDataGroup:2026mpi,
    author = "Takahashi, F. and others",
    collaboration = "Particle Data Group",
    title = "{Review of Particle Physics}",
    journal = "Int. J. Mod. Phys. A",
    volume = "41",
    number = "22",
    pages = "2630011",
    year = "2026",
    doi = "10.1142/s0217751x26300115"
}

@article{Ronchen:2012eg,
    author = "Ronchen, D. and Doring, M. and Huang, F. and Haberzettl, H. and Haidenbauer, J. and Hanhart, C. and Krewald, S. and Meissner, U. G. and Nakayama, K.",
    title = "{Coupled-channel dynamics in the reactions $\pi N \to \pi N, \eta N, K \Lambda, K \Sigma$}",
    journal = "Eur. Phys. J. A",
    volume = "49",
    pages = "44",
    year = "2013",
    archivePrefix = "arXiv",
    eprint = "1211.6998",
    primaryClass = "nucl-th"
}
\end{document}